\documentclass[twocolumn]{revtex4}
\usepackage{amsmath}
\usepackage{amsfonts}
\usepackage{amssymb}
\usepackage{graphicx}
\usepackage{pdfpages}

\providecommand{\U}[1]{\protect\rule{.1in}{.1in}}
\providecommand{\U}[1]{\protect\rule{.1in}{.1in}}
\begin{document}
\title{Generalization of Uncertainty Relation for Quantum and Stochastic Systems}
\author{T.\ Koide}
\email{tomoikoide@gmail.com,koide@if.ufrj.br}
\affiliation{Instituto de F\'{\i}sica, Universidade Federal do Rio de Janeiro, C.P. 68528,
21941-972, Rio de Janeiro, RJ, Brazil}
\author{T.\ Kodama}
\email{kodama.takeshi@gmail.com,tkodama@if.ufrj.br}
\affiliation{Instituto de F\'{\i}sica, Universidade Federal do Rio de Janeiro, C.P. 68528,
21941-972, Rio de Janeiro, RJ, Brazil}
\affiliation{Instituto de F\'{\i}sica, Universidade Federal Fluminense, 24210-346,
Niter\'{o}i, RJ, Brazil}

\begin{abstract}
The generalized uncertainty relation applicable to quantum and stochastic systems is derived within the stochastic variational method.
This relation not only
reproduces the well-known inequality in quantum mechanics but also is applicable
to the Gross-Pitaevskii equation and the Navier-Stokes-Fourier equation, showing that 
the finite minimum uncertainty between the position and the momentum is not an inherent property of quantum mechanics but a common feature of stochastic systems.
We further discuss the possible
implication of the present study in discussing the application of the hydrodynamic picture to
microscopic systems, like relativistic heavy-ion collisions.
\end{abstract}


\maketitle

\section{introduction}

Variational formulation is the standard approach to incorporate symmetries of a given system  
which play fundamental roles in its dynamics.
Unfortunately, there exist several cases where such an approach is not applicable. 
Therefore the extension of the variational principle is worthwhile to be investigated \cite{yasue1}.

Let us consider for example the variational formulation of classical mechanics.
There, the evolution of a system is described by optimizing the corresponding action with respect to virtual trajectories 
for which we can define at least the second order time derivative. 
Therefore, if we extend the domain of the virtual trajectories to include non-differentiable
trajectories such as the Brownian motion, we should introduce a new
variational approach \cite{reyes}. 
Such a generalization of the variational principle is known as the stochastic control problem in the stochastic calculus 
and 
there
are various works to generalize the variational principle in this direction
\cite{yasue1,guerra,marra,serva,jae,pav,rosen,naga,kappen,gomes,cre,eyink,arna,chen,holm,zam-rev,kkk,koide}. 
In this paper, we consider the stochastic variational method (SVM) proposed
by Yasue \cite{yasue1}.

This generalization of the variational principle provides us a possible
unified description of classical and quantum behaviors. In fact, we can derive
the Schr\"{o}dinger equation by employing the stochastic variation to the
action which leads to the Newton equation under the application of the
classical variation. Although the framework of SVM was originally proposed to
reformulate Nelson's stochastic quantization \cite{nelson}, its applicability
is not restricted to the quantization problem. The Navier-Stokes-Fourier
equation is obtain by employing the stochastic variation to the classical
action of the Euler (ideal fluid) equation \cite{kk2}. 
This method is useful also to introduce the model where the quantum and classical degrees of freedom coexist
\cite{qch}.
It is also worth mentioning that
Schr\"{o}dinger developed a classical probability theory where the probability density is given by the product of real wave functions. 
This is called reciprocal process \cite{sch31,sch32,ber32,jam74,eqm1,eqm2,yasue,vigier,zam13,leon13,koide18}.  
In Ref.\ \cite{eqm1}, it is shown that the evolution of the reciprocal process can be formulated in the form of SVM.

Such a generalized perspective enables us to find the correspondence between stochastic
and quantum behaviors. For example, there exists the well-known fundamental
limitation for simultaneous measurements 
between two canonical variables
in quantum mechanics. 
This uncertainty principle by Heisenberg constitutes one of the
intrinsic features of quantum mechanics. Mathematically, its origin traces
back to the non-commutative nature of the operators corresponding to the
observables in question. 
On the other hand, in the framework of SVM, the
observables are expressed as not only operators but also stochastic quantities
\cite{s-noether,kkk}. 
Thus the algorithm to derive the uncertainty relation in SVM is not obvious. 
Once this is clarified, we may extend it to other systems which contain 
(sometimes hidden) intrinsically stochastic nature, such as hydrodynamics.
The uncertainty relation of hydrodynamics 
may further offer a clue to find possible quantum
effects when hydrodynamic approaches are applied to microscopic systems, like
relativistic heavy-ion collisions.

As is pointed out by Nelson in Ref.\ \cite{nelson86}, when the stochastic trajectory is identified with  
the real trajectory of a particle, a presumable requirement of physics (separability) 
cannot be employed. As is seen later, however, 
the stochastic trajectory does not necessarily have a physical reality in our discussion of the uncertainty relation. 
Therefore we will not go into the details of the interpretation of the stochastic trajectory in 
Nelson's stochastic approach.

In this paper, we discuss the generalized uncertainty relation in the
framework of SVM. 
For this purpose, we introduce the Hamiltonian
formalism of SVM, which is applicable to particle systems and continuum media 
of quantum and classical dynamics on an equal footing. 
This enables us to define the
standard deviation of the momentum for general stochastic trajectories 
and hence to derive the generalized uncertainty relation. 
As a special case, this reduces to that in quantum mechanics. 
When it is applied to water at room temperature, 
we find that the obtained minimum uncertainty is two orders of magnitude 
larger than that of quantum mechanics, although it is still sufficiently small
compared to the coarse-grained scale of hydrodynamics.
Such a minimum uncertainty will play an important role in applying hydrodynamics to a microscopic system like relativistic heavy-ion collisions.

This paper is organized as follows. 
In Sec.\ \ref{sec:hamiltonian}, we first introduce the Hamiltonian form in SVM.
In Sec.\ \ref{sec:ucr}, we define the
standard deviation of the momentum in SVM and derive the inequality between
the standard deviations of the position and the momentum satisfied for
stochastic systems. The applications to continuum media is discussed in
Sec.\ \ref{sec:ucr-media}. Section \ref{sec:cr} is devoted to concluding remarks.

In the following, $k_B$ and $c$ denote the Boltzmann constant and the speed of light, respectively.

\section{Hamiltonian formulation of SVM}

\label{sec:hamiltonian}

\subsection{Stochastic Lagrangian}

Before introducing the Hamiltonian form, 
we shortly review the standard formulation of SVM. See for
example, Refs.\ \cite{zam-rev,kkk} for details.

We consider the trajectory of a (virtual) particle described 
by the following forward stochastic differential equation (SDE),
\begin{equation}
d\mathbf{r}(t)
=\mathbf{u}(\mathbf{r}(t),t)dt+\sqrt{2\nu}d\mathbf{W}_{t}~~~~(dt>0).\label{fsde}%
\end{equation}
Here $\mathbf{u}(\mathbf{x},t)$ is a field associated with the particle
velocity yet to be determined by the stochastic variation. In this paper, a
difference $dA(t)$ is always defined by $A(t+dt)-A(t)$, independently of the
sign of $dt$. The last term generates the zig-zag nature of the trajectory and called
noise term. The intensity of the noise is characterized by $\nu$. 
We consider that the noise is given by the (standard) Wiener
process $\mathbf{W}_{t}$ which satisfies the
following correlation properties,
\begin{align}
E[d\mathbf{W}_{t}]  & =0,\\
E[(dW_{t}^{i})(dW_{t}^{j})]  & =|dt|\delta^{ij},~~(i,j=x,y,z),\label{corr2}%
\end{align}
where $E[~~~]$ indicates the average over stochastic events.

The particle motion described by Eq.\ (\ref{fsde}) can be characterized also by
introducing the probability distribution defined by
\begin{align}
\rho(\mathbf{x},t) = \int d^{3} \mathbf{r}_{i}~\rho_{I} (\mathbf{r}_{i}) E[
\delta^{(3)}(\mathbf{x} - \mathbf{r}(t)) ],
\end{align}
where $\mathbf{r}(t)$ (more precisely $\mathbf{r}(t;\mathbf{r}_{i})$ with
$\mathbf{r}_{i}$ being the initial position of the particle) is the solution
of Eq.\ (\ref{fsde}) and $\rho_{I} (\mathbf{r}_{i})$ is the particle
distribution at an initial time $t_{i}$. As is well-known, the evolution
equation of $\rho(\mathbf{x},t)$ is, using Eq.\ (\ref{fsde}), given by the
Fokker-Planck equation,
\begin{equation}
\partial_{t} \rho(\mathbf{x},t) = \nabla\cdot(-\mathbf{u} (\mathbf{x},t) +
\nu\nabla) \rho(\mathbf{x},t).\label{ffp}%
\end{equation}

In the formulation of the variational method, we should fix not only an initial
condition but also a final condition. This implies that the forward SDE alone
is not sufficient. We have  to consider also a backward process in time,
$dt<0$, describing a stochastic process from the final condition to the
initial condition. That is, when the probability distribution evolves from
$\rho_{I}(\mathbf{x})$ at $t_{i}$ to $\rho_{F}(\mathbf{x})=\rho(\mathbf{x}%
,t_{f})$ at a final time $t_{f}$ following Eq.\ (\ref{ffp}), the time-reversed
process describes the evolution from $\rho_{F}(\mathbf{x})$ to $\rho
_{I}(\mathbf{x})$. Suppose that this process is given by the backward SDE,
\begin{equation}
d\mathbf{r}(t)=\tilde{\mathbf{u}}(\mathbf{r}(t),t)dt+\sqrt{2\nu}%
d\tilde{\mathbf{W}}_{t}~~~~(dt<0),\label{bsde}%
\end{equation}
where $\tilde{\mathbf{W}}_{t}$ is the Wiener process again satisfying the same
correlation properties introduced above. 
These SDEs (\ref{fsde}) and (\ref{bsde}) are relative to the increasing and decreasing sub-$\sigma$-algebras used below to define the mean forward and backward derivatives, 
respectively \cite{nelson,yasue}. 
For Eq.\ (\ref{bsde}) to describe the same statistical ensemble given by 
the forward SDE (\ref{fsde}), we find that the following consistency condition should be
satisfied \cite{kkk},
\begin{equation}
\mathbf{u}(\mathbf{x},t)=\tilde{\mathbf{u}}(\mathbf{x},t)+2\nu\nabla\ln
\rho(\mathbf{x},t).\label{cc}%
\end{equation}
The same property can be reproduced also from
the nature of the Bayesian statistics \cite{caticha}.

For the stochastic trajectories, the usual definition of the particle velocity is not applicable because
$d\mathbf{r}/dt$ is not well defined in the vanishing limit of $dt$ due to the
singular behavior of $\mathbf{W}_{t}$ (and $\tilde{\mathbf{W}}_{t}$). The
possible time differential in such a case is studied by Nelson \cite{nelson},
finding that there are two possibilities: one is the mean forward
derivative,
\begin{equation}
D \mathbf{r}(t) = \lim_{dt \rightarrow0+} E \left[  \frac{\mathbf{r}(t + dt) -
\mathbf{r}(t)}{dt} \Big| \mathcal{P}_{t} \right] ,
\end{equation}
and the other the mean backward derivative,
\begin{equation}
\tilde{D} \mathbf{r}(t) = \lim_{dt \rightarrow0-} E \left[  \frac{\mathbf{r}(t
+ dt) - \mathbf{r}(t)}{dt} \Big| \mathcal{F}_{t} \right] .
\end{equation}
These expectations are conditional averages, where $\mathcal{P}_{t}$
($\mathcal{F}_{t}$) indicates to fix values of $\mathbf{r}(t^{\prime})$ for
$t^{\prime}\le t~~(t^{\prime}\ge t)$. For the $\sigma$-algebra of all
measurable events of $\mathbf{r}(t)$, $\mathcal{P}_{t}$ and $\mathcal{F}_{t}$
represent an increasing and a decreasing family of sub-$\sigma$-algebras
\cite{cre}. Then, applying these definitions to Eqs.\ (\ref{fsde}) and
(\ref{bsde}), we obtain $D \mathbf{r}(t) = \mathbf{u}(\mathbf{r}(t),t)$ and
$\tilde{D} \mathbf{r}(t) = \tilde{\mathbf{u}}(\mathbf{r}(t),t)$, respectively.

To see how we introduce actions expressed in the above stochastic trajectory
$\mathbf{r}(t)$, let us consider the classical Lagrangian for one particle,
$L(\mathbf{r}, \dot{\mathbf{r}}) = \frac{m}{2}\dot{\mathbf{r}}^{2}(t) -
V(\mathbf{r}(t))$
where $m$ is the mass of the particle and $V(\mathbf{x})$ is a potential
distributed in $\mathbf{x}$. Due to the existence of the two definitions of
the time derivatives $D$ and $\tilde{D}$, the most general quadratic form of
the kinetic energy in terms of the stochastic trajectory is given by
\cite{koide}
\begin{eqnarray}
&&\hspace*{-0.5cm} \frac{m}{2} \dot{\mathbf{r}}^{2}(t) \longrightarrow \nonumber \\
&&\hspace*{-0.5cm} \frac{m}{2} \left[ B_{+} \{
A_{+} (D\mathbf{r}(t))^{2} + A_{-} (\tilde{D}\mathbf{r}(t))^{2} \} + B_{-}
(D\mathbf{r}(t))\cdot(\tilde{D}\mathbf{r}(t)) \right] , \nonumber \\
\end{eqnarray}
where
\begin{align}
A_{\pm}  = 1/2 \pm\alpha_{1} \ \ \ \  B_{\pm} = 1/2 \pm\alpha_{2},
\end{align}
with $\alpha_{1}$ and $\alpha_{2}$ being arbitrary real constants. These
coefficients are chosen to reproduce the classical kinetic energy in the
vanishing limit of $\nu$ where the difference between $D\mathbf{r}(t)$ and
$\tilde{D}\mathbf{r}(t)$ disappears. For related discussions, see also
Refs.\ \cite{davidson-79,hase}.

By using this expression, the stochastic action which should be optimized is
given by the expectation value of the time integral of the stochastic
Lagrangian,
\begin{align}
I[\mathbf{r}] = \int^{t_{f}}_{t_{i}} dt E \left[  L(\mathbf{r}, D\mathbf{r},
\tilde{D}\mathbf{r}) \right] .\label{sto-action}%
\end{align}
Here the stochastic Lagrangian is given by
\begin{align}
L (\mathbf{r}, D\mathbf{r}, \tilde{D}\mathbf{r})  & = \frac{m}{2} \left(
D\mathbf{r}(t), \tilde{D}\mathbf{r}(t) \right)  M \left(
\begin{array}
[c]{c}%
D\mathbf{r}(t)\\
\tilde{D}\mathbf{r}(t)
\end{array}
\right)  - V(\mathbf{r}(t)),\label{sto-lag}%
\end{align}
using
\begin{align}
M=\left(
\begin{array}
[c]{cc}%
A_{+} B_{+} & \frac{1}{2}B_{-}\\
\frac{1}{2}B_{-} & A_{-} B_{+}%
\end{array}
\right) .\label{matrixm}%
\end{align}
As will be seen later, the different values of $(\alpha_{1},\alpha_{2})$ in
$A_{\pm}$ and $B_{\pm}$ enables us to describe quantum mechanics and hydrodynamics
in a unified way.

The discussions developed so far are sufficient to implement the stochastic
variation: we obtain the stochastic Euler-Lagrange equation by applying the
stochastic variation to the stochastic action (\ref{sto-action}). To discuss
the uncertainty relation, however, we need to introduce the stochastic Hamiltonian.

Note that the particle trajectory discussed here does not necessarily have a physical reality but is 
a mathematical tool to implement our variation. 
As a matter of fact, we consider the stochastic variation for the trajectory of a fluid element in the application to continuum media, but, as is well-known, 
the fluid element is just a mathematical object which helps our understanding for the motion of fluids.

\subsection{Hamiltonian form}

Let us introduce the Legendre transformation to change the two variables
$(D\mathbf{r}(t),\tilde{D}\mathbf{r}(t))$ to $(\mathbf{p}(t),\bar
{\mathbf{p}}(t))$ in the stochastic Lagrangian,
\begin{equation}
H(\mathbf{r},\mathbf{p},\bar{\mathbf{p}})\equiv\frac{1}{2}(\mathbf{p}(t)\cdot
D\mathbf{r}(t)+\bar{\mathbf{p}}(t)\cdot\tilde{D}\mathbf{r}(t))-L(\mathbf{r}%
,D\mathbf{r},\tilde{D}\mathbf{r}). \label{sto_hami}%
\end{equation}
Here  we express ${D}\mathbf{r}$ and $\bar{D}\mathbf{r}$ in the above with $\mathbf{p}$ 
and $\bar{\bf p}$ by using the following relations, 
\begin{equation}
\mathbf{p}(t)\equiv2\frac{\partial L}{\partial(D\mathbf{r}(t))}\ \ \ \bar
{\mathbf{p}}(t)\equiv2\frac{\partial L}{\partial(\tilde{D}\mathbf{r}%
(t))}.\label{matrix_px}%
\end{equation}
The factor $1/2$ in Eq.\ (\ref{sto_hami}) (and hence the factor $2$ in
Eq.\ (\ref{matrix_px})) is introduced to reproduce the classical result in the
vanishing limit of $\nu$. This is the stochastic Hamiltonian and, after some
algebra, we find
\begin{equation}
H(\mathbf{r},\mathbf{p},\bar{\mathbf{p}})=\frac{1}{8m}\left(
\begin{array}
[c]{cc}%
\mathbf{p}(t) & \bar{\mathbf{p}}(t)
\end{array}
\right)  M^{-1}\left(
\begin{array}
[c]{c}%
\mathbf{p}(t)\\
\bar{\mathbf{p}}(t)
\end{array}
\right)  +V(\mathbf{r}(t)).\label{sto_hami2}%
\end{equation}
Here $M^{-1}$ is the inverse matrix of $M$. Note that
$\mathrm{det}~(M)\neq0$,
to have the inverse matrix.

We can construct the variational principle based on this stochastic
Hamiltonian, as is known in classical mechanics, by defining the stochastic
action which is a functional of $\mathbf{r}$, $\mathbf{p}$ and $\bar
{\mathbf{p}}$,
\begin{eqnarray}
\lefteqn{I\left[  \mathbf{r},\mathbf{p},\bar{\mathbf{p}} \right] } \nonumber \\
&=& \int_{t_{i}}^{t_{f}%
}dt E\left[  \frac{\mathbf{p}(t)\cdot D\mathbf{r}(t)+\bar{\mathbf{p}}(t)%
\cdot\tilde{D}\mathbf{r}(t)}{2}-H(\mathbf{r},\mathbf{p},\bar{\mathbf{p}%
})\right]. \label{Action_H} \nonumber \\
\end{eqnarray}
Then the variations are expressed as 
\begin{align}
\mathbf{a}(t)  & \rightarrow \mathbf{a}(t) + \delta\mathbf{h}_{\mathbf{a}%
}(\mathbf{a}(t),t) \ \ \ \ \ (\mathbf{a}=\mathbf{r},\mathbf{p},\bar
{\mathbf{p}}),
\end{align}
where $\delta\mathbf{h}_{\mathbf{a}}$ is an arbitrary smooth infinitesimal
function defined for each event, satisfying the boundary condition,
\begin{equation}
\delta\mathbf{h}_{\mathbf{a}}(\mathbf{a}(t_{i}),t_{i})=\delta\mathbf{h}%
_{\mathbf{a}}(\mathbf{a}(t_{f}),t_{f})=0.
\end{equation}

Finally, the variation of $I\left[  \mathbf{r},\mathbf{p},\bar{\mathbf{p}}
\right] $ leads to
\begin{eqnarray}
&&  D\mathbf{r}(t) = \frac{\partial H}{\partial\mathbf{p}(t)/2}, \label{eqn:mom1}\\
&&  \tilde{D}\mathbf{r}(t) = \frac{\partial H}{\partial\bar{\mathbf{p}}%
(t)/2},\label{eqn:mom2} \\
&&  \left[  \frac{\tilde{D}\mathbf{p}(t)+D\bar{\mathbf{p}}(t)}{2}+
\frac{\partial H}{\partial\mathbf{r}(t)} \right] _{\mathbf{r}(t) = \mathbf{x}}
= 0. \label{SNE}%
\end{eqnarray}
The symbol $\mathbf{r}(t) = \mathbf{x}$ in the third equation indicates that
all stochastic trajectories are replaced with the spatial parameter
$\mathbf{x}$ at the last step of the calculations. It is because we require
that the action is optimized for any distributions of the stochastic
variables. The first two equations give the definitions of $\mathbf{p}$ and
$\bar{\mathbf{p}}$ which exactly coincide with Eq.\ (\ref{matrix_px}). 
The third equation corresponds to the stochastic generalization of Newton's
equation of motion and obtained by Nelson for the first time 
without using the variational approach \cite{nelson}. 
We call this equation stochastic Newton equation.

The parameters $\alpha_{1}$ and $\alpha_{2}$ do not explicitly
appear in the stochastic Newton equation and affect only the definitions of
the momenta $\mathbf{p}$ and $\bar{\mathbf{p}}$. This fact plays an important
role in finding the standard deviation of the momentum in Sec.\ \ref{sec:ucr}.

By eliminating $\mathbf{p}$ and $\bar{\mathbf{p}}$ from Eq.\ (\ref{SNE}) using Eqs.\ (\ref{eqn:mom1}) and (\ref{eqn:mom2}), 
we recover the optimized equation in the Lagrangian formalism. In fact, by using
the mean velocity defined by
\begin{align}
\mathbf{u}_{m}(\mathbf{x},t) = \frac{\mathbf{u}(\mathbf{x},t) + \tilde
{\mathbf{u}}(\mathbf{x},t)}{2}, \label{eqn:meanvel}%
\end{align}
we can represent Eq.\ (\ref{SNE}) as 
\begin{eqnarray}
&& \left(  \partial_{t}+\mathbf{u}_{m}(\mathbf{x},t)\cdot\nabla\right)
\mathbf{u}_{m}(\mathbf{x},t) -\nabla\cdot(\mu\widehat{E}(\mathbf{x}%
,t)) \nonumber \\
&& -2\kappa\nabla\ \left(  \frac{1}{\sqrt{\rho(\mathbf{x},t)}}\nabla^{2}%
\sqrt{\rho(\mathbf{x},t)}\right)  = -\frac{1}{m} \nabla V(\mathbf{x}),
\label{NS}%
\end{eqnarray}
where
\begin{align}
\mu=\alpha_{1}(1+2\alpha_{2})\nu\ \ \ \ \kappa=2\alpha_{2}\nu^{2},
\end{align}
and $\widehat{E}\ $is a symmetric $3\times3$ tensor, 
\begin{align}
\widehat{E}^{ij} (\mathbf{x},t) = \partial_{j}{u}_{m}^{i}(\mathbf{x}%
,t)+\partial_{i}{u}_{m}^{j}(\mathbf{x},t).\label{eqn:vel-tensor}%
\end{align}
In the above, the notation $\left(  \nabla\cdot\right) $ represents the
contraction of the vector operator $\nabla$ with one of the indices of the
tensor. The dynamics of the probability distribution $\rho$ included in
Eq.\ (\ref{NS}) is given by the Fokker-Planck equation (\ref{ffp}).

In the vanishing limit of $\nu$, Eq.\ (\ref{NS}) reduces to Newton's
equation of motion. That is, SVM is the natural generalization of the classical variational approach. Depending on the choice of the parameters
$\alpha_{1}$ and $\alpha_{2}$, the optimized dynamics by Eq.\ (\ref{NS}) can
describe various different phenomena.
One of the examples is the choice of $(\alpha_{1}, \alpha_{2}) = (0,1/2)$ and
$\nu= \hbar/(2m)$, where the Schr\"{o}dinger equation is reproduced
\cite{yasue1}. To see it, we introduce the wave function by
\begin{equation}
\psi(\mathbf{x},t) = \sqrt{\rho(\mathbf{x},t)} e^{i \theta(\mathbf{x}%
,t)},\label{eqn:wf}%
\end{equation}
with $\theta$ being a velocity potential defined by
$\mathbf{u}_{m}(\mathbf{x},t) = \hbar \nabla\theta(\mathbf{x},t)/m$.
Then we find that the wave function satisfies the Schr\"{o}dinger equation
\footnote{Exactly speaking, it is considered that one more condition for the
velocity $\mathbf{u}_{m}$ should be employed to reproduce exactly the
Schr\"{o}dinger equation \cite{taka,wall,derak}.},
\begin{equation}
i \hbar\partial\psi(\mathbf{x},t) = \left[  -\frac{\hbar^{2}}{2m} \nabla^{2} +
V(\mathbf{x}) \right]  \psi(\mathbf{x},t).
\end{equation}
That is, the quantization of classical systems can be regarded as the
stochastic optimization of classical actions.

Our Hamiltonian formulation developed here contains two momenta $\mathbf{p}$
and $\bar{\mathbf{p}}$, but this is different from the
Hamiltonian formulation developed in Ref.\ \cite{misawa-hamil}.  
There the Lagrangian which is symmetric for $D$ and $\tilde{D}$ is employed and 
thus only one conjugate momentum associated with the mean velocity $\mathbf{u}_{m} (\mathbf{x},t)$ is
considered. Note however 
that the introduction of the two momenta presented here plays a crucial
role in the construction of the generalized framework of the uncertainty relation.

\subsection{Continuum medium}

The idea developed so far is easily extended to continuum media \cite{kk2}.
Let us consider the motion of the fluid which has a mass density $\rho_{m}
(\mathbf{x},t)$. In the classical variation, there are at least two approaches
to describe the dynamics of continuum media; one is in the Euler coordinate
and the other in the Lagrange coordinate. To apply SVM, it is convenient to
use the latter.

First, we decompose the fluid into small pieces called fluid elements. The
trajectory of the fluid element which is initially located at $\mathbf{R}$ is
denoted by $\mathbf{r}(\mathbf{R},t)$. Then the Lagrangian of this continuum
medium in the Lagrange coordinate is given by \cite{kk2,hugo}
\begin{eqnarray}
L  &=& \int d^{3}\mathbf{R}\rho_{m}^{0}(\mathbf{R}) \nonumber \\
&& \times \left[  \frac{1}{2}\left(
\frac{d\mathbf{r}(\mathbf{R},t)}{dt}\right)  ^{2}-\frac{1}{m}V(\mathbf{r}%
(\mathbf{R},t)) -\frac{\varepsilon(\rho_{m}(\mathbf{r}(\mathbf{R},t),t))}%
{\rho_{m}(\mathbf{r}(\mathbf{R},t),t)}\right]  \nonumber\\
&=& \int d^{3}\mathbf{R}\ \frac{\rho_{m}^{0}(\mathbf{R})}{m}\mathcal{L}%
(\mathbf{r},\dot{\mathbf{r}}),\label{eqn:cl-med}%
\end{eqnarray}
where $V(\mathbf{x})$ and $\varepsilon(\rho_{m}(\mathbf{x},t))$ represent the
external potential and the internal energy density of the fluid, respectively.
The latter is a differentiable function of the mass density $\rho
_{m}(\mathbf{x},t)$. The integral is taken over the initial position $\mathbf{R}$ of
the fluid elements with the initial mass distribution $\rho_{m}^{0}%
(\mathbf{R})=\rho_{m}(\mathbf{R},t_{i})$. To define the Lagrangian function $\mathcal{L}$, 
we introduce $m$ as the averaged mass per unit constituent particle which agrees
with the mass of the constituent particle itself in the case of the simple
(one-component) fluid. For the classical Lagrangian in the Euler
coordinate, see, for example, Ref.\ \cite{elze}.

The initial and posterior mass densities $\rho_{m}^{0}(\mathbf{R})$ and
$\rho_{m}(\mathbf{r}(\mathbf{R},t),t)$ are given by the relation in the
absence of the turbulence,
$\rho_{m}(\mathbf{r}(\mathbf{R},t),t)=\rho_{m}^{0}(\mathbf{R})/J$,
where $J$ is the Jacobian,
$J=\mathrm{det}\left\vert \partial\mathbf{r}(\mathbf{R},t)/
\partial\mathbf{R}\right\vert $.
See Refs.\ \cite{kk2,hugo} for details.

In the application of SVM, we consider the position of the fluid element
$\mathbf{r}(\mathbf{R},t)$ as a stochastic variable, which satisfies
Eqs.\ (\ref{fsde}) and (\ref{bsde}). Replacing the trajectory of the fluid
element with the corresponding stochastic variable, the stochastic Lagrangian
function is obtained,
\begin{eqnarray}
\mathcal{L}(\mathbf{r},D\mathbf{r},\tilde{D}\mathbf{r})  
&=& m\left[  \frac
{1}{2}(D\mathbf{r}(\mathbf{R},t),\tilde{D}\mathbf{r}(\mathbf{R},t))M\left(
\begin{array}
[c]{c}%
D\mathbf{r}(\mathbf{R},t)\\
\tilde{D}\mathbf{r}(\mathbf{R},t)
\end{array}
\right)  \right. \nonumber \\
&& \left. -\frac{1}{m}V(\mathbf{r}(\mathbf{R},t))-\frac{\varepsilon(\rho
_{m}(\mathbf{r}(\mathbf{R},t)))}{\rho_{m}(\mathbf{r}(\mathbf{R},t),t)}\right], \label{eqn:lag-cm}
\end{eqnarray}
where $M$ is the matrix defined by Eq.\ (\ref{matrixm}).

Then the momenta are defined through the Legendre
transformation as
\begin{equation}
\mathbf{p}(\mathbf{R},t)=2\frac{\partial\mathcal{L}}{\partial(D\mathbf{r}%
)} \ \ \ \ \bar{\mathbf{p}}(\mathbf{R},t)=2\frac{\partial\mathcal{L}}%
{\partial(\tilde{D}\mathbf{r})}.
\end{equation}
These definitions still coincide with those in the particle case, noting that
$\rho_{m}^{0}(\mathbf{R})=m\delta^{(3)}(\mathbf{R}-\mathbf{r}(t_{i}))$ in such
a case.
Thus the stochastic action is 
\begin{eqnarray}
\lefteqn{I[\mathbf{r},\mathbf{p},\bar{\mathbf{p}}]=} && \nonumber \\
&& \int_{t_{i}}^{t_{f}}dt \int d^{3}\mathbf{R}\frac{\rho_{m}%
^{0}(\mathbf{R})}{m}E\left[  
\frac{{\bf p}\cdot D{\bf r}+\bar{\bf p}\cdot \tilde{D}{\bf r}}{2}
- 
\mathcal{H}(\mathbf{r},\mathbf{p},\bar{\bf p})\right],\nonumber \\
\end{eqnarray}
where
${\cal H}({\bf r},{\bf p},\bar{\bf p})
= \frac{1}{2}({\bf p}\cdot D{\bf r}+\bar{\bf p}\cdot \tilde{D}{\bf r}) - 
{\cal L}(\mathbf{r},D\mathbf{r},\tilde{D}\mathbf{r})$.

Implementing the stochastic variation, 
we find
\begin{equation}
 \left(
\begin{array}
[c]{c}%
\mathbf{p}(\mathbf{R},t)\\
\bar{\mathbf{p}}(\mathbf{R},t)
\end{array}
\right)  =2mM\left(
\begin{array}
[c]{c}%
\mathbf{u}(\mathbf{r}(\mathbf{R},t),t),
\tilde{\mathbf{u}}(\mathbf{r}(\mathbf{R},t),t)
\end{array}
\right) ,
\end{equation}
and
\begin{align}
& \left[  \frac{\tilde{D}\mathbf{p}+D\bar{\mathbf{p}}}{2}+\nabla V(\mathbf{r})\right.  \nonumber\\
& \left.  -\frac{m}{\rho_{m}(\mathbf{r},t))}\ \nabla\left(
\frac{d}{d(1/\rho_{m}(\mathbf{r},t)))}\frac{\varepsilon(\rho
_{m}(\mathbf{r},t))}{\rho_{m}(\mathbf{r},t))}\right)  \right]  _{\mathbf{r}(\mathbf{R},t)=\mathbf{x}}%
=0.\label{SNE2}%
\end{align}
The last equation is, 
by introducing the mean velocity as Eq.\ (\ref{eqn:meanvel}), expressed as
\begin{eqnarray}
&& \left(  \partial_{t}+\mathbf{u}_{m}(\mathbf{x},t)\cdot\nabla\right)
\mathbf{u}_{m}(\mathbf{x},t)  -\frac{\nabla\cdot
(\mu\rho_{m}(\mathbf{x},t)\widehat{E}(\mathbf{x},t))}{\rho_{m}(\mathbf{x},t)}\nonumber\\
&& -2\kappa\nabla\ \left(  \frac{1}{\sqrt{\rho_{m}(\mathbf{x},t)}}\nabla
^{2}\sqrt{\rho_{m}(\mathbf{x},t)}\right)  \nonumber \\
&& =-\frac{1}{m}\nabla V(\mathbf{x})
-\frac{\nabla P(\mathbf{x},t)}{\rho_{m}(\mathbf{x},t)},\label{eqn:eq-vari-cont}%
\end{eqnarray}
where the definitions of $\mu$, $\kappa$ and the symmetric $3\times3$ tensor
$\widehat{E}\ $ are the same as before, and $P$ is the adiabatic pressure
defined as
\begin{equation}
P(\mathbf{x},t)=\rho_{m}^{2}(\mathbf{x},t)\frac{d}{d\rho_{m}(\mathbf{x}%
,t)}\frac{\varepsilon(\rho_{m}(\mathbf{x},t))}{\rho_{m}(\mathbf{x},t)},
\end{equation}
which satisfies the thermodynamic relation. See also Ref.\ \cite{kk2} for
technical details.

In the vanishing limit of $\nu$ where $\mu= \kappa= 0$,
Eq.\ (\ref{eqn:eq-vari-cont}) is reduced to the Euler equation which describes
the dynamics of the ideal fluid.
It is known that
Eq.\ (\ref{eqn:eq-vari-cont}) describes the Gross-Pitaevskii and
Navier-Stokes-Fourier equations by choosing the parameters appropriately. See
Ref.\ \cite{kk2} and the discussion in Sec.\ \ref{sec:ucr-media}.

\section{Uncertainty relation}

\label{sec:ucr}

\begin{figure}[t]
\begin{center}
\includegraphics[scale=0.2]{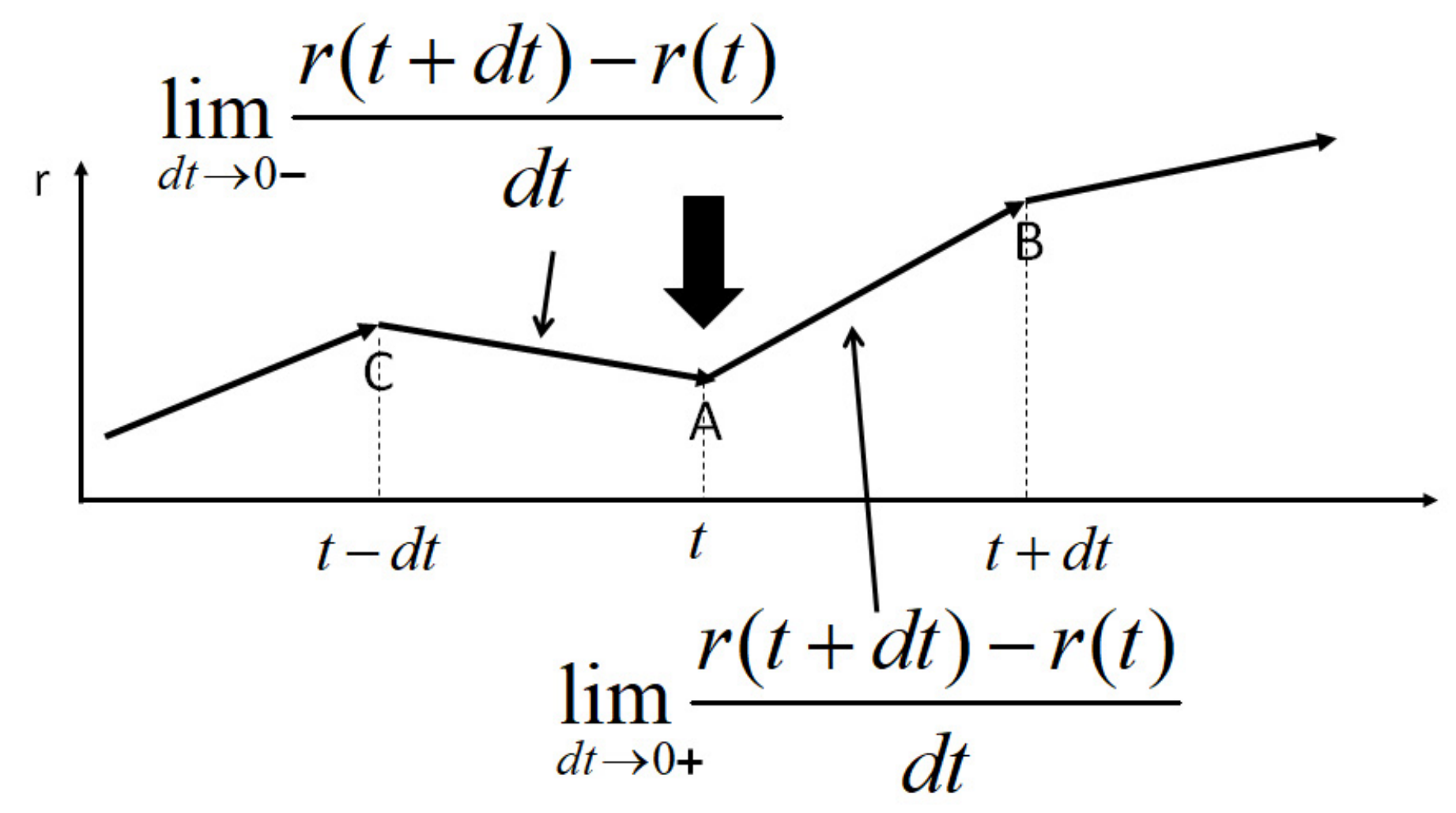}
\end{center}
\caption{The schematic figure for a stochastic trajectory and velocities.
There are at least two different possibilities for the velocity at the point
A.}%
\label{def-vel}%
\end{figure}

In this section, we derive the generalized uncertainty relation for stochastic
systems described in SVM. The definitions introduced below are applicable to
both of the particle system and the continuum medium.

For the sake of the unified description, we introduce the notation for the
average with respect to the stochastic events and the mass distribution as
\begin{align}
\left\{  A \right\} _{av} = \left\{
\begin{array}
[c]{ccl}%
E[A] &  & \mathrm{for\ the\ particle\ system }\\
\int d^{3} \mathbf{R}\ \frac{\rho^{0}_{m} (\mathbf{R})}{m} E[A] &  &
\mathrm{for\ the\ continuum\ medium }.
\end{array}
\right.
\end{align}

With this, it is natural to define the standard deviation of the position by
\begin{equation}
\Delta_{x^{i}} = \frac{1}{N} \sqrt{\left\{  (r^{i}- \left\{  r^{i} \right\}
_{av} )^{2}\right\} _{av}}, \label{unc_x}%
\end{equation}
with the normalization factor,
\begin{align}
N = \left\{
\begin{array}
[c]{ccl}%
1 &  & \mathrm{for\ the\ particle\ system}\\
\int d^{3} \mathbf{R}\ \frac{\rho^{0}_{m} (\mathbf{R})}{m} &  &
\mathrm{for\ the\ continuum\ medium }. \label{norma}%
\end{array}
\right.
\end{align}
Thus, this represents the standard deviation of the position
for the particle and that of the center of mass for the continuum medium, respectively.

On the other hand, because of the two different quantities $\mathbf{p}$ and
$\bar{\mathbf{p}}$, the definition of the standard deviation for the momentum
is not straightforward in SVM. The appearance of the two momenta is associated
with the two possible definitions of the velocities which come from 
the
lack of the differentiability 
of the stochastic trajectories as is shown in
Fig.\ \ref{def-vel}: there are two different inclinations of the particle 
trajectory at the point A, ${\displaystyle \lim_{dt \rightarrow 0_+ }({r}(t+dt) -
{r}(t))/dt}$ and ${\displaystyle \lim_{dt \rightarrow 0_+ } ({r}(t-dt) -
{r}(t))/(-dt)}$. These two limits generally do not coincide in stochastic systems.
See also the discussion for the uncertainty principle in the original paper by Heisenberg  \cite{heisenberg}.

As is seen from the stochastic Newton equations (\ref{SNE}) and (\ref{SNE2}),
the two momenta contribute the evolutions of the particle and of the fluid
element on an equal footing, independently of the choice of the parameters
$\alpha_{1}$ and $\alpha_{2}$. Therefore, it will be adequate to define the
standard deviation of the momentum by the symmetric average of the two
standard deviations for $\mathbf{p}$ and $\bar{\mathbf{p}}$ as
\begin{equation}
\Delta_{p^{i}}\equiv\sqrt{\frac{\left\{  (p^{i}-\left\{  p^{i} \right\}
_{av})^{2}\right\} _{av} +\left\{  (\bar{p}^{i} -\left\{  \bar{p}^{i} \right\}
_{av})^{2}\right\} _{av}}{2}} . \label{delta_p}%
\end{equation}
The above definitions reproduce the standard
deviation in quantum mechanics, as is shown in Appendix \ref{app1}.

To calculate the standard deviation of the momentum, we introduce the sum and
the subtraction of the two momenta,
\begin{equation}
\mathbf{p}_{+}  \equiv\frac{\mathbf{p}+\bar{\mathbf{p}}}{2} \ \ \ \ \
\mathbf{p}_{-}  \equiv\frac{\mathbf{p}-\bar{\mathbf{p}}}{2},
\end{equation}
respectively. Moreover, using the Cauchy-Schwarz inequality, we can show that
$\{ {A}^{2} \}_{av} \{ {B}^{2} \}_{av} \geq|\{ {AB} \}_{av}|^{2}$ 
for real stochastic variables
${A}$ and ${B}$. Then we find the following inequalities,
\begin{align}
& \Delta_{x^{i}}\Delta_{p_{+}^{j}}\geq\frac{m}{N}\left\vert \left\{
(r^{i}-\left\{  r^{i}\right\}  _{av})(u_{m}^{j}-\left\{  u_{m}^{j}\right\}
)\right\}  _{av}-\mu N\delta_{ij}\right\vert ,\\
& \Delta_{x^{i}}\Delta_{p_{-}^{j}}\geq\frac{m}{\nu N}\left\vert \mu\left\{
(r^{i}-\left\{  r^{i}\right\}  _{av})(u_{m}^{j}-\left\{  u_{m}^{j}\right\}
_{av})\right\}  _{av}-\kappa N\delta_{ij}\right\vert ,
\end{align}
where
$\Delta_{{p}_{\pm}^{j}}=\sqrt{\left\{  ({p}_{\pm}^{j}-\left\{  {p}_{\pm}%
^{j}\right\}  _{av})^{2}\right\}  _{av}}$.
The standard deviation of the momentum defined above can be
expressed as
$\Delta_{p^{i}}=\sqrt{\Delta_{p^{i}_+}^{2}+\Delta_{p^{i}_{-}}^{2}}$.
Therefore, the product of the standard deviations of the position and the
momentum satisfies the following inequality,
\begin{align}
\lefteqn{\Delta_{x^{i}}\Delta_{p^{j}}}\nonumber\\
& \geq\frac{m}{N}\sqrt{\left(  1+\left(  \frac{\mu}{\nu}\right)  ^{2}\right)
(A^{ij}) ^{2}+N^{2}\frac{(\kappa-\mu^{2})^{2}}{\nu
^{2}+\mu^{2}}\delta_{ij}} \label{rs-ucr} \\
& \geq m\nu\chi\delta_{ij},\label{ucr}%
\end{align}
where
\begin{eqnarray}
A^{ij} &=& \left\{  (r^{i}-\left\{  r^{i}\right\}  _{av})(u_{m}^{j}-\left\{
u_{m}^{j}\right\}  _{av})\right\}  _{av} \nonumber \\
&& -\delta_{ij}\frac{\mu N(\kappa+\nu^{2})}{\nu^{2}+\mu^{2}},  \\
\chi &=& \frac{4}{\sqrt{1+(\mu/\nu)^{2}}}|\mathrm{det}~(M)|,
\end{eqnarray}
and $\mathrm{det}~(M)$ is calculated from Eq.\ (\ref{matrixm}) as
\begin{equation}
\mathrm{det}~(M)=\frac{\kappa-\mu^{2}}{4\nu^{2}}.\label{det}%
\end{equation}
This inequality is our main result and represents the generalized uncertainty
relation satisfied for quantum and stochastic systems formulated in SVM.

As was mentioned, we need to set $(\alpha_{1},\alpha_{2},\nu) =
(0,\ 1/2,\ \hbar/(2m))$ to reproduce the Schr\"{o}dinger equation. Then the
above standard deviation of the momentum agree with the well-known expression
in quantum mechanics,
$\Delta_{p^{i}} = \sqrt{\langle(\hat{p}^{i} - \langle\hat{p}^{i} \rangle)^{2}
\rangle}$
where $\hat{p}^{i} = -i \hbar\partial_{x^{i}}$ and $\langle\ \ \rangle$
denotes the expectation value with a wave function. See Appendix\ \ref{app1}
for details. Therefore Eq.\ (\ref{ucr}) reproduces the uncertainty relation in
quantum mechanics called Kennard inequality,
$\Delta_{x^{i}}\Delta_{p^{j}}\geq\frac{\hbar}{2}\delta_{ij}$.

One should remember that the optimization in SVM reproduces the result of the
classical variation in vanishing limit of $\nu$. Therefore one can easily see the trivial result 
that the finite minimum uncertainty disappears in classical systems. On the
contrary, the finiteness of the minimum uncertainty is held for any stochastic
dynamics with a finite $\nu$ because $\mathrm{det}~(M) \neq0$ for any
$(\alpha_{1},\alpha_{2})$.

\subsection{Robertson-Schr\"{o}dinger uncertainty relation}

As a more general uncertainty relation, there exists the Robertson-Schr\"{o}dinger uncertainty relation.
When this uncertainty relation is applied to the position and momentum operators, 
we find the following inequality,
\begin{eqnarray}
\Delta^2_{x^i} \Delta^2_{p^j} 
&\ge& \frac{\hbar^2}{4} + \left| \frac{1}{2} \langle [ \hat{x}^i, \hat{p}^j ]_+ \rangle  - \langle \hat{x}^i \rangle \langle \hat{p}^j \rangle \right|^2 \nonumber \\
&=& \frac{\hbar^2}{4} + ( {\rm Re}[ \langle (\hat{x}^i-\langle \hat{x}^i \rangle)(\hat{p}^j-\langle \hat{p}^j \rangle) ] )^2, 
\end{eqnarray}
where $[~]_+$ denotes the anti-commutator.

On the other hand, when we employ the parameters to reproduce the quantum mechanical results in SVM.
the same result can be obtained from Eq.\ (\ref{rs-ucr}), noting that 
\begin{eqnarray}
\lefteqn{\left\{  (r^{i}-\left\{  r^{i}\right\}  _{av})(u_{m}^{j}-\left\{
u_{m}^{j}\right\}  _{av})\right\}  _{av} } && \nonumber \\
&=& \frac{1}{m}{\rm Re}[ \langle (\hat{x}^i-\langle \hat{x}^i \rangle)(\hat{p}^j-\langle \hat{p}^j \rangle) ]. 
\end{eqnarray}
Therefore the Robertson-Schr\"{o}dinger uncertainty relation is obtained when Eq.\ (\ref{rs-ucr}) is used while 
the Kennerd inequality is derived for Eq.\ (\ref{ucr}).

\section{Application to continuum media}

\label{sec:ucr-media}

So far, two examples in the continuum media are known to be described in the
framework of SVM: one is the Gross-Pitaevskii equation and the other the
Navier-Stokes-Fourier equation. We study our uncertainty relations in these cases.

\subsection{Gross-Pitaevskii equation}

In the usual derivation of the Gross-Pitaevskii equation, we start from the
bosonic system described by the following Hamiltonian operator,
\begin{eqnarray}
\hat{H} 
&=& \int d^{3} \mathbf{x} \left[  \frac{\hbar^{2}}{2m} \nabla\hat{\phi
}^{\dagger}(\mathbf{x}) \cdot\nabla\hat{\phi}(\mathbf{x}) + V(\mathbf{x})
\hat{\phi}^{\dagger}(\mathbf{x}) \hat{\phi}(\mathbf{x}) \right. \nonumber \\
&& \left. + \frac{U_{0}}{2}
(\hat{\phi}^{\dagger}(\mathbf{x}) \hat{\phi}(\mathbf{x}))^{2} \right]
,\label{eqn:qft-lag}%
\end{eqnarray}
where $m$, $V(\mathbf{x})$ and $U_{0}$ are the mass of an atom, an external
potential and the coupling constant of the two-body interaction,
respectively.
The bosonic field operator satisfies
\begin{align}
[\hat{\phi}(\mathbf{x},t),\hat{\phi}^{\dagger}(\mathbf{x}^{\prime},t)] 
=\delta^{(3)}(\mathbf{x}-\mathbf{x}^{\prime}).\label{eqn:qft-com}%
\end{align}
By applying the mean-field approximation to this quantum-field theoretical
system, we obtain the Gross-Pitaevskii equation which describes the behavior
of the Bose-Einstein condensate of the ultracold atom.

The same equation can be obtained in a different way. 
Let us consider a classical gas composed of $N$ particles. To describe the
exact behavior of the gas, we need to solve the $N$-body Newton equations. On
the other hand, when we are interested in only macroscopic behaviors of the
gas, we do not need to know the positions and velocities of the particles.
Instead, it is sufficient to observe the behaviors of the reduced number of
macroscopic variables: the mass density and the velocity field. Then the
many-body particle system is regarded as a continuum medium. 
The approximated quantum dynamics of such a many-body system can be obtained by
applying SVM to the Lagrangian of the continuum medium \cite{kk2}.

Let us consider the situation in the ultracold atom, choosing 
the internal energy density in Eq.\ (\ref{eqn:cl-med}) as
\begin{equation}
\varepsilon(\rho_{m}(\mathbf{x},t))=\frac{U_{0}}{2}\frac{\rho^{2}_m (\mathbf{x},t)}{m^2}.
\end{equation}
Other parameters are the same as those in
the case of the Schr\"{o}dinger equation; $(\alpha_{1},\alpha_{2}%
,\nu)=(0,\ 1/2,\ \hbar/(2m))$. Substituting these into Eqs.\ (\ref{ffp}) and
(\ref{eqn:eq-vari-cont}) and defining the wave function $\psi(\mathbf{x},t)$
in the same way as Eq.\ (\ref{eqn:wf}), 
the Gross-Pitaevskii equation is reproduced,
\begin{equation}
i\hbar\partial_{t}\psi(\mathbf{x},t)=\left[  -\frac{\hbar^{2}}{2m}\nabla
^{2}+V(\mathbf{x})+U_{0}|\psi(\mathbf{x},t)|^{2}\right]  \psi(\mathbf{x},t),
\end{equation}
See Ref.\ \cite{kk2} for more details.

Now, by using Eq.\ (\ref{ucr}), the uncertainty relation obtained from our
formula is expressed as
\begin{equation}
\Delta_{x^{i}}\Delta_{p^{j}}\geq\frac{\hbar}{2}\delta_{ij}.
\end{equation}
This inequality is the same as the one obtained in the quantum field theory
from Eq.\ (\ref{eqn:qft-com}), where the standard deviations are represented
by
\begin{align}
\Delta_{x^{i}}  & =\frac{1}{\langle\int d^{3}\mathbf{x}\ \hat{\phi}^{\dagger
}(\mathbf{x},t)\hat{\phi}(\mathbf{x},t)\rangle}\sqrt{\langle(\hat{x}^{i}%
)^{2}\rangle-\langle\hat{x}^{i}\rangle^{2}},\label{delx}\\
\Delta_{p^{i}}  & =\sqrt{\langle(\hat{p}^{i})^{2}\rangle-\langle\hat{p}%
^{i}\rangle^{2}},\label{delp} \\
\hat{\mathbf{x}} &  =\int d^{3}\mathbf{x}\hat{\phi}^{\dagger}\mathbf{x}\hat{\phi}, \\
\hat{\mathbf{p}}  & =\int d^{3}\mathbf{x}\hat{\phi}^{\dagger}(-i\hbar\nabla)\hat{\phi},
\end{align}
where $\langle\ \ \rangle$ denotes the expectation value with a state
vector in a Fock space. That is, our uncertainty relation is consistent with
that of the quantum field theory.

As a different derivation of the Gross-Pitaevskii equation in SVM, see
Ref.\ \cite{loffred}. The bosonic field quantization in the framework of SVM
is studied in Ref.\ \cite{svm-field}.

\subsection{Navier-Stokes-Fourier equation}

So far we have exclusively considered the case where the noise term is
interpreted as the origin of the quantum fluctuations where $\nu=\hbar/(2m)$.
Now we consider a classical stochastic system where $\nu$ characterizes the
intensity of the thermal fluctuations and thus is given by a function of
temperature. We further choose the parameters $\alpha_{2}=0$, then $\kappa$
vanishes and our optimized equation (\ref{eqn:eq-vari-cont}) is reduced to
\begin{equation}
\rho_{m}\left(  \partial_{t}+\mathbf{u}_{m}\cdot\nabla\right)  \mathbf{u}%
_{m}-\nabla\cdot(\eta\widehat{E})=-\nabla P. \label{eqn:nsf}%
\end{equation}
Here we take $V=0$ for the sake of simplicity. Then
$\eta\widehat{E}$ represents the stress tensor and this is the
Navier-Stokes-Fourier equation \footnote{Here the second coefficient of
viscosity is not reproduced. For this, we have to consider the entropy
dependence in the internal energy density. See Refs.\ \cite{kk2} for details.}.
We introduced the coefficient of viscosity $\eta=\mu\rho_{m}$. Then
$\mu=\alpha_{1}\nu=\eta/\rho_{m}$ is known as kinematic viscosity.

In the modern perspective in statistical mechanics, the macroscopic
dissipation (viscosity) is induced by the fluctuation of the underlying
microscopic degrees of freedom. This property results in the
fluctuation-dissipation theorem. We can consider that the SVM formulation of the
Navier-Stokes-Fourier equation naturally reflects this aspect. Moreover
the coefficient of viscosity is formally expressed with the velocity (current)
correlation function of the constituent particles,
\begin{equation}
\mu=\frac{\alpha_{1}}{3N}\int_{-\infty}^{\infty}dt\theta(t)\left\{
\delta\mathbf{v}(\mathbf{r},t)\cdot\delta\mathbf{v}(\mathbf{r},0)\right\}
_{av},
\end{equation}
with $\delta\mathbf{v}(\mathbf{r},t)=d\mathbf{r}(\mathbf{R},t)/dt-\mathbf{u}%
(\mathbf{r}(\mathbf{R},t))$.

Substituting these parameters into our inequality (\ref{ucr}), the uncertainty
relation for hydrodynamics is given by
\begin{eqnarray}
\Delta_{x^{i}}\Delta_{p^{j}}\geq\frac{m\mu^{2}}{\sqrt{\mu^{2}+\nu^{2}}}%
\delta_{ij}. \label{eqn:ucr-fluid}
\end{eqnarray}
Note that the standard deviations can be expressed as the averages with
respect to the mass density distribution because, for an arbitrary function $F(\mathbf{x},t)$,
$\left\{  F(\mathbf{r}(\mathbf{R},t),t)\right\}  _{av}=\int d^{3}
\mathbf{x}\ \frac{\rho_{m}(\mathbf{x},t)}{m}F(\mathbf{x},t)$.

Equation (\ref{eqn:ucr-fluid}) means that the minimum uncertainty becomes
larger as the increase of the viscosity of the fluid. This is reasonable,
because due to the fluctuation-dissipation theorem,  larger fluctuation
induces higher viscosity and vice-versa. 

The dimension of $\mu$ and $\nu$ is $L^{2}T^{-1}$ and, for a macroscopic fluid, the
magnitude of $\mu$ which represents the viscosity of fluids, 
is of the order of the macroscopic scale. On the other
hand, $\nu$ is the noise intensity in the microscopic dynamics described by
Eq.\ (\ref{fsde}).
Then the magnitude of the associated distance ($L$) is the
microscopic scale, while that of $LT^{-1}$ should be, at a maximum, of the order of the sound velocity. 
Thus the magnitude of $\nu$ is expected to be
much smaller than that of $\mu$.  Let us estimate the minimum
uncertainty, for example, of water. Assuming the simple fluid, $m$ is
equivalent to the mass of the water molecule, $\sim3\times10^{-26}$ kg. The
kinematic viscosity at room temperature is $\sim1\times10^{-6}$ $m^{2}/s$.
Then the minimum uncertainty is estimated as
\begin{equation}
\frac{m\mu^{2}}{\sqrt{\mu^{2}+\nu^{2}}}\sim m\mu\sim3\times10^{-32}%
\ [kg\ m^{2}/s^{2}]\sim100\times\frac{\hbar}{2}.
\end{equation}
This value is two orders larger than that of quantum mechanics
but is still much smaller than the (coarse-grained) macroscopic scales of
hydrodynamics. 
To observe this uncertainty for a fluid, we need to measure the mass
distribution and the velocity field of fluids with precision. In such a
precision, however, the hydrodynamic description normally looses its
validity because the hydrodynamic approach is applicable only to macroscopic
variables observed in a coarse-grained scale which will be much larger than
$\hbar$.
See also the discussion in Sec.\ \ref{sec:cr}.

\subsubsection{The possible role of the $\kappa$ term in hydrodynamics}

In the discussion above, we set $\alpha_{2} = 0$ to ignore the coefficient
$\kappa= 2 \alpha_{2} \nu$ because such a term is usually considered to be
irrelevant to the dynamics of fluids, while the same term, which is called quantum potential,
plays an important role in the hydrodynamic representation of the
Sch\"{o}dinger equation \cite{holland}
(See the last term on the left hand side of Eq.\ (\ref{NS})).
Recently, Brenner pointed out that, 
since the velocity of a tracer particle of fluids is not necessarily parallel
to the mass velocity, the existence of these two velocities should be taken
into account in the formulation of hydrodynamics. 
By applying the linear irreversible thermodynamics, then, it is found that 
the difference of the two velocities are characterized by the mass density gradient  
as is our self-consistency condition (\ref{cc}) and 
a new term analogous to the quantum potential can appear. 
This modified theory is called the bivelocity hydrodynamics
\cite{brenner1,brenner2}. 
The similar results are obtained in various
approaches including SVM, but the existence is still controversial
\cite{klimontovich,ottinger,graur,greensh,eu,don,dadzie,kk2,gustavo}. 
See also TABLE 1 of Ref.\ \cite{brenner2}. It is also worth mentioning that the
structure which is the same as the bivelocity hydrodynamics naturally appears
as the next-to-leading order relativistic corrections to the
Navier-Stokes-Fourier equation \cite{gustavo}.

Another possible role of the $\kappa$ term is related to the sign of the
kinetic term in the stochastic Lagrangian. To obtain the Navier-Stokes-Fourier
equation, we chose $\alpha_{1} > 0$ and $\alpha_{2} = 0$, leading to
\begin{equation}
\mathrm{Tr}~(M)>0~\mathrm{and}~\det~(M)<0,
\end{equation}
where $M$ is the two-by-two matrix (\ref{matrixm}) appearing in the stochastic Lagrangian. On
the other hand, from the well-known second-order algebra, the necessary and
sufficient condition for the positive-semidefinite kinetic term in
Eq.\ (\ref{sto_hami2}) (or equivalently Eq.\ (\ref{eqn:lag-cm})) is given by
\begin{equation}
\mathrm{Tr}~(M)>0~\mathrm{and}~\det~(M)>0,
\end{equation}
and it demands
\begin{equation}
\mu<\sqrt{\kappa}. \label{mu<k}%
\end{equation}
Therefore, to have a finite viscosity requiring the positive kinetic term,
$\kappa$ should be finite. Note, however, that it is not clear whether such a
requirement is mandatory, because the above positivity is irrelevant to the
positivity of the fluid energy,
$\int d^{3} \mathbf{x} \left[  \frac{1}{2}\rho_{m}(\mathbf{x},t) \mathbf{u}%
^{2}_{m} (\mathbf{x},t) + \varepsilon(\rho_{m}(\mathbf{x},t)) \right] $
which is positive independently of the value of $\kappa$.

This $\kappa$ term may play an important role in applying the hydrodynamic model to microscopic systems such as 
relativistic heavy-ion collisions, as is discussed in Sec.\ \ref{sec:cr}.

\section{Concluding remarks}

\label{sec:cr}

We discussed the uncertainty relation satisfied for general stochastic systems
which are described in the stochastic variational method. We first developed
the Hamiltonian formulation of SVM. Then we introduced the two
different momenta through the Legendre transformation of the stochastic
Lagrangian. Because the contributions from the two momenta are the same in the
stochastic Newton equation, the standard deviation of the momentum should be
defined by the symmetric average of the standard deviations of those momenta.
Using this, we derived the generalized uncertainty relation satisfied for
stochastic systems described in SVM. When we choose the parameter set which
reproduces the Sch\"{o}dinger equation, our uncertainty relation is reduced to
the Kennard inequality in quantum mechanics.
Note that the SVM quantization does not always reproduce the same result as the canonical
quantization. As an example, see Appendix \ref{app2}.

The stochastic generalization of the variational principle 
indicates that we need to use the two optimizations
quite differently, depending on the hierarchy of observation scales in
physics. Suppose that we obtain an equation by optimizing an action. When we
observe a system with a macroscopic scale where we can ignore the possible
zig-zag motion, it is sufficient to use classical variation for
the optimization. However, when we are interested in more
precise behaviors in a microscopic scale and hence we cannot ignore the
non-differentiable trajectories in the optimization of the same action, 
the stochastic variation should be employed.

In this paper, we have focused on the uncertainty relation of the position and the momentum. 
To obtain the uncertainty relation of, for example, the angle and the angular momentum, 
we have to express Eq.\ (\ref{fsde}) in the spherical coordinate but then 
the transformation law of the standard Wiener process ${\bf W}_t$ is non-trivial. See, for example, Ref. \cite{gardinar}. 
The generalization of the present formulation for such a case is left as a future work.

Applying the present approach to the continuum media, we investigated the
uncertainty relation for the Gross-Pitaevskii equation and the
Navier-Stokes-Fourier equation. For the former, we confirmed that our
result is the same as that obtained in the quantum field theory. For the
latter, we found that the minimum uncertainty is enhanced as the increase
of the viscosity. This is a reasonable behavior because,  from the fluctuation-dissipation theorem,  
higher viscosity 
indicates larger fluctuation and vice-versa.
For the water at room temperature, the minimum uncertainty becomes
two orders of magnitude larger than that of quantum mechanics, $\hbar/2$. 
It is
however much smaller than the (coarse-grained) macroscopic scales of
hydrodynamics and thus the effect of the minimum uncertainty is negligible in
the usual fluids. 
In our formulation, the minimum uncertainty never
disappear for any stochastic systems, independently of the values of
$(\mu,\kappa)$ when the momenta are well-defined. 
Therefore, we can consider 
that the finite minimum uncertainty between the position and the momentum is not an inherent feature in quantum mechanics 
but a common feature in stochastic systems described in SVM.

However, the existence of the minimum uncertainty for continuum media shows that a special care should be taken when a hydrodynamic approach is applied to microscopic systems.
For example, the hydrodynamic
models are often employed to describe the behaviors of the hot matters created
in relativistic heavy-ion collisions \cite{review-kodama}. Although the
uncertainty relation for the relativistic fluid is not obtained here, 
it is reasonable to assume that the minimum value cannot be smaller than the quantum
mechanical one, $\hbar/2$. 
The global size of the initially produced fluid is the
order of 10 fm, but the distribution of the energy density is expected to be
inhomogeneous and presents many local peaks \cite{review-kodama}. 
To describe such a
peak as a fluid, the uncertainty of the position of the fluid elements must be
smaller than the size of the peaks, say, $10^{-1}$ fm. Therefore the expected
uncertainty of the momentum becomes, at least, $\sim1$ GeV/c. 
If we attribute this fluctuation to the pure thermal origin without any quantum effect, then the corresponding
temperature would be much
larger than $\sim1$ GeV/$k_{B}$ $\sim10^{13}$ K. Such a temperature far
exceeds what we usually expect in the relativistic heavy-ion collisions. 
However, as is seen from the form of the $\kappa$ term (quantum potential),
the effect of the quantum fluctuation is extremely enhanced near the large
inhomogeneities. Therefore, if we employ a relativistic hydrodynamic approach
to describe the above situation, it will be necessary to consider a model which incorporates dynamically 
the enhanced quantum effect near the inhomogeneity.
Then in the presence of a sharp peak, the generated pressure should be attributed
not only to a high temperature of the hot matter, but also to the quantum
fluctuation. 
Such effects are expected to be important to shed a new light to understand 
the collective dynamics observed in small systems as proton-nucleon and proton-proton collisions.

\begin{acknowledgements}
The authors acknowledge the financial support by CNPq (307516/2015-6), 
FAPERJ, CAPES, PRONEX. 
A part of the work was developed under the project INCT-FNA Proc.\ No.\ 464898/2014-5. Kodama thanks
to the hospitality extended by H.\ St\"{o}cker, D.\ Rischke and M.\ Bleicher, during his stay at FIAS.
\end{acknowledgements}

\appendix

\section{Relation to quantum mechanical standard deviation}

\label{app1}

We will show that Eq.\ (\ref{delta_p}) is equivalent to that in quantum
mechanics. See also the discussion in Ref.\ \cite{falco}. First we find the
second order correlation of the momentum operator in quantum mechanics as
\begin{eqnarray}
\langle\hat{\mathbf{p}}^{2} \rangle 
&=& \langle\hbar^{2}
(\nabla R(\mathbf{x},t))^{2} + (\nabla S(\mathbf{x},t))^{2} \rangle,
\end{eqnarray}
where $\langle~~~\rangle$ represents the expectation value with the wave
function $\psi(\mathbf{x},t)= e^{R(\mathbf{x},t)-iS(\mathbf{x},t)/\hbar}$ with
the real functions $R(\mathbf{x},t)$ and $S(\mathbf{x},t)$.
On the other hand, Eq.\ (\ref{cc}) is re-expressed with this decomposition of
the wave function as
$\mathbf{u}(\mathbf{x},t) - \tilde{\mathbf{u}}(\mathbf{x},t) = 2\hbar\nabla R(\mathbf{x},t) /m$.
Note that the current of the probability density is equivalent to that of the
Fokker-Planck equation. Therefore the mean velocity is expressed as
$\mathbf{u}_{m} (\mathbf{x},t) 
= \nabla S/m
(\mathbf{x},t)$.

Using these relations, we can show that
$ E[ (m\mathbf{u})^{2} + (m\tilde{\mathbf{u}})^{2} ]/2 = \langle
\hat{\mathbf{p}}^{2} \rangle$
and
$E[m\mathbf{u}] = E[m \tilde{\mathbf{u}}] = \langle\nabla S
\rangle/m= \langle\hat{\mathbf{p}} \rangle$.
Summarizing these results, we find
\begin{eqnarray}
&& \hspace*{-1cm} (\Delta_{\mathbf{p}})^{2} 
= \langle\hat{\mathbf{p}}^{2} \rangle- \langle
\hat{\mathbf{p}} \rangle^{2}  \nonumber \\
&& \hspace*{-1cm}= \frac{E[(m\mathbf{u})^{2}] - (E[m\mathbf{u}%
])^{2}}{2} + \frac{E[(m\tilde{\mathbf{u}})^{2}] - (E[m\tilde{\mathbf{u}}%
])^{2}}{2}.
\end{eqnarray}
For the case of quantum mechanics where $(\alpha_{1},\alpha_{2}) = (0,1/2)$
and $\nu= \hbar/(2m)$, we have $\mathbf{p} = m \mathbf{u}$ and $\bar
{\mathbf{p}} = m \tilde{\mathbf{u}}$. Therefore the right hand side of the
above result is equivalent to Eq.\ (\ref{delta_p}).

\section{Uncertainty relation for quantum dissipative system}

\label{app2}

There is an example where SVM leads to the different result from the canonical
quantization \cite{misawa,skag}. We consider the following time-dependent
Lagrangian,
\begin{align}
L({r},\dot{r},t)=e^{\gamma t}\left[  \frac{m}{2}\left(  \frac{d{r}(t)}%
{dt}\right)  ^{2}-V({r}(t))\right] , \label{cla_diss_ac}%
\end{align}
where $\gamma$ is a constant which characterizes the time scale of the
relaxation of a particle with the mass $m$. It is easily confirmed that we
obtain a dissipative equation from this Lagrangian by applying the classical
variational method.

The canonical quantization of this system was studied by Caldirola
\cite{caldi} and Kanai \cite{kanai}, leading to the following Schr\"{o}dinger
equation,
\begin{align}
i\hbar\psi = \left[  -\frac{\hbar^{2}}{2m}e^{-\gamma t}\partial^{2}_{x}
+ e^{\gamma t}V({x}) \right]  \psi,
\end{align}
defining the momentum operator $\hat{ p} = -i\hbar\partial_{x}$. Therefore,
the corresponding uncertainty relation is
\begin{align}
\Delta_{x} \Delta_{p} \ge\frac{\hbar}{2}.\label{eqn:ucr-ck}%
\end{align}

The SVM quantization of the same Lagrangian was studied in Ref.\ \cite{misawa}. 
The stochastic Lagrangian is then expressed as
\begin{equation}
L({r},D{r},\tilde{D}{r},t)=e^{\gamma t}\left[  m\frac{(D{r}(t))^{2}+(\tilde
{D}{r}(t))^{2}}{4}-V({r}(t))\right]  .
\end{equation}
We consider that ${r}(t)$ satisfies the same stochastic differential equations
(\ref{fsde}) and (\ref{bsde}), and $\nu= \hbar/(2m)$. Then the stochastic
variation leads to
\begin{equation}
i\hbar\partial_{t} \psi = \left[  - \frac{\hbar^{2}}{2m} \partial
^{2}_{x} + V({x}) + \frac{i\hbar}{2}\gamma\ln\frac{\psi^{*} }{\psi} + a(t) \right]  \psi ,
\end{equation}
where $a(t)$ is an arbitrary time-dependent factor, which can be absorbed into
the definition of the phase \cite{misawa}. This non-linear Schr\"{o}dinger
equation coincides with the so-called Kostin (or Schr\"{o}dinger-Langevin)
equation \cite{kostin,dekker,hasse} by choosing
$a(t) = -\frac{i\hbar}{2}\gamma\left\langle \ln (\psi^{*}/\psi) \right\rangle$
where $\langle~~\rangle$ means the expectation value with $\psi$.

In the original approach by Kostin, this equation is derived
phenomenologically to satisfy Eherenfest's theorem and hence the definition of
the momentum operator is not clear. On the other hand, applying the method
developed in this paper, we find
\begin{align}
\Delta_{x} \Delta_{p} \ge\frac{\hbar}{2} e^{\gamma t},\label{ucr-kostin}%
\end{align}
where the momentum operator is obtained by the stochastic Noether
theorem \cite{kkk}, finding $\hat{p} = - i\hbar e^{\gamma t} \partial_{x}$.

As is seen from the right hand side, the minimum uncertainty increases with
time. This behavior is different from the Caldirola-Kanai theory, but in
reality, the behavior of Eq.\ (\ref{eqn:ucr-ck}) is considered to be
problematic. Following the discussion in
Refs.\ \cite{dekker,hasse,brittin,havas,sent}, the uncertainty of the inertia
is defined by $\Delta_{inertia} = e^{-\gamma t} \Delta_{p}$ and then
Eq.\ (\ref{eqn:ucr-ck}) leads to the following minimum uncertainty of the
inertia,
\begin{align}
\Delta_{x} \Delta_{inertia} \ge\frac{\hbar}{2} e^{-\gamma t}.
\end{align}
It shows that the uncertainty vanishes in the asymptotic limit in time and
this behavior is sometimes considered to be unphysical
\cite{dekker,hasse,brittin,havas,sent}. On the other hand, such a behavior
does not occur in our result (\ref{ucr-kostin}).

It should be, however, noted that the above violation of the uncertainty
relation is still controversial. In Ref.\ \cite{dieter}, it is argued that if
we change the definition of the momentum operators by multiplying or dividing
a factor $e^{\gamma t}$ as was done to define the uncertainty of the inertia,
we need to change also the Hilbert space. If this possibility is taken into
account, the minimum uncertainty for the inertia in the Caldirola-Kanai theory
does not vanish asymptotically.

There is no established method for the systematic
quantization of dissipative systems and the aforementioned theories are only
one part of various approaches. As other approaches, see, for example,
Refs.\ \cite{dieter2,goldin,marcelo,blasone} and references therein.

\end{document}